\documentclass[fleqn,usenatbib]{mnras}

\usepackage{fix-cm}
\usepackage{newtxtext,newtxmath}
\usepackage[T1]{fontenc}

\DeclareRobustCommand{\VAN}[3]{#2}
\let\VANthebibliography\thebibliography
\def\thebibliography{\DeclareRobustCommand{\VAN}[3]{##3}\VANthebibliography}

\usepackage{graphicx}	% Including figure files
\usepackage{amsmath}	% Advanced maths commands

\title[high-frequency polarisation properties of MACS J0717.5$+$3745]{Crossing the veil of the brightest radio relic in the sky MACS J0717.6$+$3745 }

\author[name]{
A. Pasetto,$^{1}$\thanks{E-mail: a.pasetto@irya.unam.mx}
}
\author[A. Pasetto et al.]{
A. Pasetto,$^{1}$\thanks{E-mail: a.pasetto@irya.unam.mx}
O. Vilchis,$^{1}$
A. Bonafede,$^{2,3}$
E. F. Jim\'enez-Andrade,$^{1}$
K. Rajpurohit,$^{4}$
E. Murphy$^{5}$
\\
$^{1}$Instituto de Radioastronom\'ia y Astrof\'isica, Universidad Nacional Aut\'onoma de M\'exico, Antigua Carretera a P\'atzcuaro, 8701, 58089 Morelia,  M\'exico\\
$^{2}$Dipartimento di Fisica e Astronomia, Universit\'a di Bologna, Via P. Gobetti 93/2, 40129 Bologna, Italy\\
$^{3}$ INAF-Istituto di Radio Astronomia, Via Gobetti 101, 40129 Bologna, Italy\\
$^{4}$ Center for Astrophysics | Harvard \& Smithsonian, 60 Garden Street, Cambridge, MA 02138, USA\\
$^{5}$ National Radio Astronomy Observatory, 520 Edgemont Road, Charlottesville, VA 22903, USA\\
}

\date{Accepted XXX. Received YYY; in original form ZZZ}

\pubyear{2025}

\begin{document}
\label{firstpage}
\pagerange{\pageref{firstpage}--\pageref{lastpage}}
\maketitle

% Abstract of the paper
\begin{abstract}

We present high-frequency, full-polarisation Jansky Very Large Array (VLA) radio data at X-band of the radio relic: MACS J0717.5+3745. Radio relics trace shock waves in the intracluster medium (ICM) produced during mergers. Understanding the physical characteristics of relics is important for determining their nature, whether for example they are thermal ICM electrons that are accelerated, or whether they are fossil electrons re-accelerated by a merger event. Radio spectropolarimetric analysis, such as the Stokes QU-fitting, provides a diagnostic of the nature and structure of the magnetized plasma internal or external to the source, with important implications for theoretical models. The high-frequency polarisation analysis presented here shows, for the first time, a change in the magneto-ionic structure compared to the low-frequency data available in the literature. These high-frequency, polarised data could be interpreted also with an internal depolarisation behaviour and this new finding may be used to investigate possible particle acceleration mechanism. If that is true, the change in the behaviour of the polarised signal could be tracing physical properties of a population of non-thermal particles that are undergoing to a re-acceleration of particles in the relic by large-scale internal shocks of Active Galactic Nuclei jet fossil particles ejected from the central Narrow Angle Tail radio galaxy. New upcoming broad-band VLA X- and Ku-bands data will clarify this.
Finally, we conclude that high-frequency, high-sensitive, spectropolarimetric radio data should be explored further, as they can effectively trace shock fronts and thereby provide insights into the intrinsic magneto-ionic properties of radio components.
\end{abstract}

% Select between one and six entries from the list of approved keywords.
% Don't make up new ones.

%%%%%%%%%%%%%%%%%%%%%%%%%%%%%%%%%%%%%%%%%%%%%%%%%%

%%%%%%%%%%%%%%%%% BODY OF PAPER %%%%%%%%%%%%%%%%%%
 %%me sirve para futures referencias\citet{Fournier1901},

\section{Introduction}
The intracluster medium (ICM) in galaxy clusters hosts a non-thermal component of cosmic-ray electrons (CRe) that emit synchrotron radiation in the presence of weak magnetic fields (i.e. of the order of $\sim\mu$G). Radio haloes are generally associated with non-thermal radio emission at the centres of galaxy clusters, whereas radio relics are typically located at the peripheries of merging clusters and are believed to result from the (re)-acceleration of electrons by shock events propagating through the ICM. Both structures are extended, Mpc-scale objects ($\sim$0.5–2.0 Mpc); however, relics are generally characterised by elongated morphologies and significant linear polarisation, typically up to $\sim$20–30\% \citep[see reviews by][]{2014IJMPD..2330007B, vanWeeren2019}. According to \citet{Kempner2004}, large-scale radio sources in galaxy clusters can be subdivided into three categories: giant radio relics, relics of active galactic nuclei (AGNs), and radio phoenixes.

\textit{i)} Giant radio relics are large, arc-shaped regions of diffuse radio emission found at the peripheries of merging galaxy cluster, thought to be synchrotron emissions of electrons accelerated directly from thermal plasma in merger/acceleration shocks, or by fossil electrons that AGN or other radio galaxy activity, have previously accelerated \citep[e.g.,][]{Pinzke2013,vanWeeren2017}. 

\textit{ii)} Relics of AGNs are linked to extinct or dying AGNs, where the AGN that originated the lobes is no longer active, and no re-acceleration or compression has taken place. These relics are typically found close to the dominant galaxy in the cluster, i.e., near the first central kpc of the cluster.

\textit{iii)} Radio phoenixes are related to AGNs but also depend on the interaction with the ICM. Initially, the plasma is ejected from the AGN and dispersed into the surrounding medium, which eventually ages so that it no longer emits synchrotron radiation at observable radio frequencies. However, when a shock from a cluster merger passes through this aged plasma (faded relic), it compresses the fossil plasma and re-energizes the electrons to energies that make them radio visible again \citep[e.g.,][]{Ensslin2001,EnsslinBruggen2002,vanWeeren2011,deGasperin2015}.

Some physical properties of radio relics are well established, such as their radio spectrum ($S_{\nu} \propto \nu^{-\alpha}$) showing a very steep spectral index. Moreover, a correlation between the integrated spectral index and their morphology has also been found with $\alpha_{\scriptscriptstyle{0.3 \, \text{GHz}}}^{\scriptscriptstyle{1.4 \, \text{GHz}}}\, \simeq[$ 1 -- 1.6] for elongated radio relics and $\alpha_{\scriptscriptstyle{0.3 \, \text{GHz}}}^{\scriptscriptstyle{1.4 \, \text{GHz}}} \, \simeq[$ 1.1 -- 2.9] for roundish radio relics \citep{Feretti2012}. The analysis of the spectral index distribution gives information about the age of the synchrotron structures and shock acceleration mechanism, which are supposed to take place in these objects \citep{Stroe2014, Botteon2020}.

Nevertheless, another important feature for properly understanding the formation of these large structures is the analysis of the strength and configuration of the magnetic field using the radio polarisation information. Radio relics are important tracers of the magnetic field at the peripheries of the clusters. According to cosmological cluster simulations, magnetic field strength at the centres of galaxy clusters is predicted to be at the level of $\mu$G, decreasing in strength with distance from the centre \citep[e.g.,][]{Bonafede2010,Vazza2018,Nelson2024}. This behaviour has also been reported in previous observational studies \citep[e.g.,][]{Bonafede2009,vanWeeren2012,pearce2017vla,Rajpurohit2022}. Since most of the radio relics are at the outskirts of the clusters where the magnetic field should be very weak, the polarisation analysis of these objects can help us trace the structure of the magnetic field at the cluster's periphery \citep{van2009diffuse}. In fact, the radio polarisation properties, such as fractional polarisation ($f_{p}$) and rotation measures (RM) are excellent diagnostics of ICM physics \citep{Ensslin1998, Govoni2006, Pasetto2021review}. Up to date, several studies have investigated radio relics using full polarisation and low-frequency radio data \citep[e.g.,][]{Bonafede2009,vanWeeren2012,DiGennaro2021,Rajpurohit2021, Rajpurohit2022, Stuardi2022}. Among these, the galaxy cluster MACS J0717.5$+$3745 relic has been a significant focus for exploring polarisation properties. For example, \cite{Bonafede2009} explored the cluster's magnetic field structure through low-frequency polarised emissions, while \cite{Rajpurohit2022} conducted a comprehensive multi-band study (L, S, and C bands), analyzing variations in Rotation Measure (RM) and the dispersion of the Rotation Measure ($\sigma_{\text{RM}}$) across the relic.
Theoretical studies on shock simulations in galaxy clusters suggest the need for an old population of AGN electrons as seeds for a radio relic to be visible \citep[e.g.,][]{DominguezFernandez2024}. However, the precise relationship between the mechanisms of electron acceleration and the properties of the magnetic fields remains unclear.

This study provides a comprehensive polarisation analysis conducted using the Jansky Very Large Array (VLA) at X-band (10 GHz), focusing on the merging galaxy cluster MACS J0717.5$+$3745 (J2000 RA: \(07^{\text{h}}17^{\text{m}}31.22^{\text{s}}\) and DEC: \(+37^{\circ}45'22.6''\)), which is a massive \citep[M$_{500}= 2.52
\pm 0.12$~$\times$~$10^{14} \, $M$_\odot$,][]{2016A&A...590A.126A} merging cluster at a redshift of $z= 0.5458$ \citep{edge2003discovery}, for which, using the cosmological values of H$_{0}$ = 67.8 km s$^{-1}$ Mpc$^{-1}$, $\Omega_{matter}$ = 0.308, $\Omega_{vacuum}$ = 0.692, the angular scale of 1$\arcsec$ translates into 6.57 kpc. It is one of the most complex dynamically disturbed systems \citep[e.g., ][]{Ebeling2001, edge2003discovery}. Optical and X-ray observations reveal that the cluster consists of at least four merging subclusters \citep[][]{Limousin2016, vanWeeren2017-835}. Its X-ray luminosity is L$_{X, 0.1-2.4 \ keV}$= 2.4 $\times$ 10$^{45}$ erg s$^{-1}$ placing it as one of the hottest clusters known with an ICM temperature of T= 12.2 $\pm$ 0.4 keV \citep[][]{Ebeling2007, vanWeeren2017-835}. It hosts an extended radio halo of approximately 2.2 Mpc \citep[][]{Rajpurohit2021-646} with a radio power at 1.4 GHz of $5 \times 10^{25}$ W Hz$^{-1}$ \citep{van2009diffuse}, being the most powerful halo known to date. The cluster shows a complex morphology (see Fig.~\ref{fig:relicWB}a), with an elongated radio relic source at its centre, along which six regions are identified, such as the northern region (NR), regions R1-to-R4 \citep[regions where shock fronts are supposed to be present, see ][]{van2009diffuse}, and the Narrow-Angle-Tail (NAT) radio galaxy. 

Among the vast number of radio halos and relics in the sky
(i.e.> 100), there exist radio observations above 10 GHz only for two of them : CIZA J2242.8+5301 ("Sausage" cluster) and the Toothbrush relic \citep[][]{Stroe2014-445,Kierdorf2017,Loi2017}. For the first time, thanks to the high performance capabilities of the VLA, we detect emission at 10 GHz for both the continuum and the polarisation along the MACS J0717.5$+$3745 radio relic. A sensitivity of $\approx$5 $\mu$Jy beam$^{-1}$ and $\approx$3 $\mu$Jy beam$^{-1}$ (1$\sigma$ RMS) has been reached for the continuum and the polarised intensity data, respectively.  

In Section \S\ref{sec:2} we present the observations and data reduction, and Section \S\ref{sec:3} shows the depolarisation modelling. The results are shown in Section \S\ref{sec:4} and the discussion is presented in Section \S\ref{sec:5}. Finally, the summary and conclusion are given in Section \S\ref{sec:6}.

\section{Radio observations and data reduction}
\label{sec:2} 
 
Observations were performed with the VLA of the National Radio Astronomy Observatory (NRAO)\footnote{The National Radio Astronomy Observatory is a facility of the National Science Foundation operated under cooperative agreement by Associated Universities, Inc.} in a single session on May the 31st, 2020 (NRAO Project Code: 20A-527) using the X band receiver at C configuration in full polarisation mode. The standard continuum mode (2 MHz width channels) covering the full available frequency range, 8-12 GHz of the X-band receivers, has been used. The calibrator 3C147 has been used for the bandpass and flux density calibration, while 3C138 was adopted to calibrate the polarisation angle. To determine the D-terms, the primary unpolarised calibrator J0713+4349 was used. The target of study, MACSJ0717.5$+$3745 has been observed for a total of $\approx$ 1hr on-source. 

The data were calibrated using the Common Astronomy Software Applications (CASA) package (version 6.2.1.7). We used the latest available version of the NRAO pipeline for VLA continuum data, but modified to include polarisation calibration after the complex gain calibration. A total of 3.7\%  of the data were flagged due to radio frequency interference.

For the polarisation calibration, we followed a procedure similar to that used in \cite{Pasetto2018,Pasetto2021}. Essentially, we used a polarised model for 3C138, taking into account the changes of the Stokes Q and U parameters within the observed frequency band \cite{Perley2013}. Then, we obtained solutions for the polarisation for each 2 MHz spectral channel. As a first step, the cross-hand delays have been corrected using the flux calibrator 3C147. Subsequently, the D-terms were corrected using the unpolarised calibrator. Finally, to set the absolute electric vector polarisation angle (EVPA), the polarised model of 3C138 was used. The calibrated target needed self-calibration. During this step, models were obtained from natural weighted images, using the \emph{tclean-scale} option to account for the extended structures of the relic. We performed several cycles of imaging and self-calibration (phase) until the quality (RMS noise and SNR) of the resulting maps do not show any further improvements. 

The X-band images presented in this work were produced using the task \emph{tclean} from CASA. When using the available 4 GHz wide-band data, the multi-scale multi-frequency synthesis \citep[MS-MFS; see][]{Rau2011} with nterms=2 option, was used. The spectrum of each synchrotron component is modeled by a Taylor expansion about the reference frequency, $\nu_0$=10~GHz. The parameter nterms=2 in \emph{tclean} results in an expansion of first order. This is equivalent to assuming a spectral power-law variation of the flux ($S_\nu$) with frequency ($\nu$) in the form $S_\nu \propto \nu^\alpha$, where $\alpha$ is an average spectral within the used bandwidth. This is the expected form of the spectrum for an optically thin synchrotron emission. The final total band Stokes I, Q and U images were produced using a robust=1 weighted and primary beam corrected, resulting in images with an angular resolution of $\theta$=2$\farcs${967}$\times$2$\farcs${244}. The final continuum image of the radio relic, shown in Fig. \ref{fig:relicWB}a, reaches an rms noise of $\sim$ 5$\mu$Jy beam$^{-1}$. The polarised intensity map, shown in figure Fig.~\ref{fig:relicWB}b, reaches an rms noise level of $\sim$ 3$\mu$Jy beam$^{-1}$. 

The continuum and polarisation intensity maps at S- and C-bands available in the literature \citep[][]{Rajpurohit2022} have also been used for the spectral index and the spectropolarimetric analysis. For observation details, we refer to \citep{van2016discovery, vanWeeren2017-835}. The S- and C-bands maps were produced combining the VLA D, C and B configurations and convolved at a circular beam with a resolution of $\theta$=5$\arcsec$. 

In order to study the behaviour of the continuum and polarisation properties of the source among the broad-band frequency range from S- to X-band, the new 10 GHz Stokes I, Q and U maps needed a correction on the angular resolution. The X-band, high angular resolution Stokes I, Q and U images have been used for the high-frequency continuum and polarisation description, while for a proper broad-band (i.e. from S- to X- bands) spectral index, curvature and spectropolarimetric analysis, the 10 GHz Stokes I, Q and U images were convolved to the same angular resolution of the archival low-frequency data of $\theta$=5$\arcsec$. The minimum baseline at all the bands consider in this work, allow us to detect synchrotron emission at scales which are larger than the real angular size of the source of study. In fact, the source of study is $\approx$ 95 \arcsec large and the values of the large angular scales at the three frequencies, S-, C- and X-bands, are 490\arcsec, 240\arcsec and 145\arcsec, respectively. Therefore, the source of study is fully contained within the S-, C-, and X-band radio maps.

\section{Depolarisation modelling}
\label{sec:3} 

When a radio source emits synchrotron radiation, part of this radiation is linearly polarised, which means that the electric field oscillates in a preferred direction; this linear polarisation emission can be described in terms of the Stokes parameters for the total intensity, I, and the orthogonal components, Q and U, as follows
\citep[see ][]{Burn1966, sokoloff1998depolarization}:

\begin{equation}
    P(\lambda^{2}) =  Q(\lambda^{2}) + iU(\lambda^{2})= p(\lambda^{2}) \cdot I(\lambda^{2}) \cdot \exp\left[2i \, \chi(\lambda^{2})\right],
    \label{e4.1}
\end{equation}

\noindent where $p(\lambda^{2})$ represents the fractional polarisation, $ \chi(\lambda^{2})$ is the polarisation angle, and $\lambda$ is the observing wavelength; these quantities are wavelength-dependent and can be written as follows: 
\begin{equation}
    p(\lambda^{2}) = \frac{ P(\lambda^{2})}{I(\lambda^{2})} = \frac{\sqrt{Q^2(\lambda^{2}) + U^2(\lambda^{2})}}{I(\lambda^{2})},
    \label{e4.2}
\end{equation}
and 

\begin{equation}
\chi(\lambda^{2}) = \frac{1}{2} \arctan\left({\frac{U(\lambda^{2})}{Q(\lambda^{2})}}\right).
\label{e4.3}
\end{equation}

When polarised radiation passes through a magneto-ionic medium such as the ICM, the plane of the polarised wave is rotated by the Faraday effect as follows:

\begin{equation}
\chi(\lambda^{2}) = \chi_{0} + \text{RM}\lambda^{2},
\label{e4.4}
\end{equation}

\noindent where $\chi_{0}$ corresponds to the intrinsic polarisation angle at the source and $\text{RM}$ is the Rotation Measure. The Faraday RM is one of the most important effect to consider when carrying out polarimetric radio observations. In the case of a homogeneous magnetized medium, the RM is defined as: 
$\mu$
\begin{equation}
\frac{\text{RM}}{\text{rad} \,\text{m}^{-2} } = 0.81 \int_{source}^{observer} B_{\parallel}\times  n_{e}\times dl,
\label{e4.5}
\end{equation}

\noindent where B$_{\parallel}$ [$\mu$G] is the magnetic field component along the line of sight, $n_{e}$ [cm$^{-3}$] is the thermal electron density, and \textit{l} [pc] is the path length. One way to obtain the RM value is to perform a linear fit to equation \eqrefpar{e4.4}, therefore, at least three measurements at three different frequencies are required. However, once the RM is determined, it is not easy to extract direct information about the three physical values, B$_{\parallel}$, $n_{e}$ and \textit{l}, unless some of these values can be obtained from other observations.
Another way to obtain the polarisation properties of a source is to model fit the Stokes parameters Q and U, using analytical models representing the depolarisation mechanisms: the so-called Stokes QU-fitting\citep[see ][]{o2012complex,Pasetto2021review}. In the simplest case, the equation \eqrefpar{e4.1} can be written as:

\begin{equation}
    P(\lambda^{2}) = p_{0} \cdot \exp\left[2i \left(\chi_{0} + \text{RM} \, \lambda^2\right)\right].
    \label{e4.6}
\end{equation}

\noindent In equation \eqrefpar{e4.6} the polarised emission remains constant throughout the spectrum and a linear behaviour of $\chi$ with $\lambda^2$, is predicted. This behaviour is true only when the polarised emission from the synchrotron source passes through a physically disconnected and homogeneous magneto-ionic plasma, e.g. a simple external Faraday screen. However, real cases are much more complicated and result in complex variations of the polarisation angle and the fractional polarisation. In the literature, there are several theoretical descriptions of more realistic cases that consider the rotating and depolarizing material to be either external or internal to the synchrotron-emitting source with the magnetic field ordered and/or turbulent. All of these scenarios predict a decrease in the fractional polarisation at longer wavelengths as a consequence of the Faraday rotation. These models can be gathered under two big families: the external and internal Faraday depolarisation models \citep[for more details see ][]{Burn1966, sokoloff1998depolarization}. 

According to \cite{Burn1966}, an external Faraday depolarisation occurs when the magneto-ionic medium is only rotating and not emitting synchrotron radiation. This can be described as: 

\begin{equation}
    P(\lambda^2) = p_{0} \,I\, \exp\left(-2 \sigma_{\text{RM} }^2 \lambda^4 \right) \, \exp\left[2i \left(\chi_{0} + \text{RM}  \lambda^2 \right)  \right],
    \label{e4.7}
\end{equation}

\noindent where $\sigma_{\text{RM} }$ is defined as the dispersion over the mean RM across the source in the sky.

On the other hand, the internal Faraday depolarisation is a phenomenon that occurs when both the emitting and the rotating regions are mixed and this system contains a turbulent and regular magnetic field. Such internal Faraday depolarisation can be described as:

\begin{equation}
 P(\lambda^2)= p_{0}\, I \,\exp\left(2i \chi_{0} \right) \left(\frac{1- \exp{\left(-S\right)}}{S}\right) , 
    \label{e4.8}
\end{equation}

\noindent where S= 2$\sigma_{RM}^2$$\lambda^4$--2$i$ $\phi$$\lambda^2$. In this case, the depolarisation occurs because of the combination of both the presence of a
regular magnetic field and the turbulent magnetic field. In this equation $\sigma_{RM}$ is the internal Faraday dispersion of the random field and $\phi$ is the general description of the Faraday depth through the region. When $\sigma_{RM}$=0 (i.e., no turbulent magnetic field component), the emitting and rotating regions are only co-spatial in the presence of a regular magnetic field. Therefore, the complex degree of polarisation is given by \citep[see][]{Burn1966, sokoloff1998depolarization}: 

\begin{equation}
P(\lambda)=p_0I\exp [2i(\chi_0+RM\lambda^2)]\left(\frac{{\rm sin} (2\Delta RM\lambda^2)}{2\Delta RM\lambda^2}\right),
\label{Burn66}
\end{equation}

\noindent where $\Delta_{\text{RM}}$ represents both a linear gradient in RM across the emission region, or internal Faraday rotation in a uniform field.

The above mentioned depolarisation equations have been used for a proper analysis of the polarisation properties of the source under study using the Stokes QU-fitting. By applying this technique, we could study the structure of the magnetized plasma in an attempt to describe the physics experienced by the source. In addition, by performing this modelling, we could obtain physical constraints of the magnetized sources directly from the polarisation data \citep{Pasetto2021review}. 

To perform the Stokes QU-fitting on the X band data, four 1GHz-wide Stokes I, Q, and U images were used for this purpose and all four images were convolved to the larger beam of $\theta$ = 3$\farcs${500}$\times$2$\farcs${642} to ensure proper comparison. Moreover, to increase the data points for the Stokes QU model fit the VLA S- and C-bands Stokes I, Q, and U images from the already published work by \cite{Rajpurohit2022}, have been used. This enables the performance of a spectropolarimetric analysis considering the wider frequency coverage from 2 to 12 GHz, after convolving all the images to the largest angular resolution of that work being $\theta$= 5$\farcs${}. Before performing the depolarisation fitting, the values of the Stokes Q and U parameters were normalized with the corresponding Stokes I value to eliminate the effects that the variations of the intensity could cause. The quantities q($\lambda^{2})$ = Q/I and u($\lambda^{2}$)= U/I have been finally used for the final depolarisation analysis \citep[as also reported in previous works e.g.,][]{o2012complex,DiGennaro2021,Pasetto2021review}.

The Markov Chain Monte Carlo (MCMC) method\footnote{MCMC is an algorithm
designed and used for numerical problems with a complex or high-dimensional distribution since it allows for efficient exploration
of the entire space and the generation of representative samples. This method is often more efficient than conventional numerical methods. For
more details, see \cite{hastings1970monte}.} has been used to model fit. It helps to determine the best parameters for the models: simple Faraday (\ref{e4.6}), an external (\ref{e4.7}), and an internal with linear RM gradient (\ref{Burn66}) Faraday depolarisation. The parameters of the equations \eqrefpar{e4.7} and \eqrefpar{Burn66}  ( $p_{0}$, $\chi_{0}$, $\text{RM}$, $\sigma_{\text{RM}}$, and $\Delta_{\text{RM}}$) were constrained to physical conditions as follows:

\[ \begin{cases} \begin{array}{lcl} 
10^{-5} & \leqslant \hspace{0.1in} p_{0} & \leqslant 0.75 \\ 
0 & \leqslant \hspace{0.1in}  \chi_{0} & < \pi \\ -200
& < \hspace{0.1in}  \text{RM} & < 200 \\ 
0 & < \hspace{0.1in} \sigma_{\text{RM}} & \leqslant 200 \\
0&< \hspace{0.1in} \Delta_{\text{RM}} & \leqslant 200 
\end{array} \end{cases} \]

\noindent The limits for the parameters were assumed based on their possible values; for example, the fractional polarisation is set up to 75$\%$, and the angle value is set to a maximum of $\pi$ because the polarisation vectors do not have a preferred direction, i.e., the values of 0 and $\pi/2$ represent the north/south and east/west directions. The $\text{RM}$ value is set to a range of $\pm$ 200 to scan a reasonable space of values for such object, and finally, $\sigma_{\text{RM}}$ and $\Delta{\text{RM}}$ are set to such range in the same way. 

\noindent
The code performed MCMC only when the signal-to-noise ratio (SNR) criterion was $>3\sigma$ for the total intensity I and the polarised flux density P data, and only when this condition was true for at least four sub-band images.
The intervals of solutions showing MCMC acceptance rates below 20\% or exhibiting parameter degeneracies were systematically excluded from the final analysis. We derived the parameter estimates for accepted intervals from the 16th, 50th, and 84th percentiles of the marginalized posterior distributions, providing robust uncertainty quantification.

\section{Results}
\label{sec:4}
\subsection{Wide-band continuum and polarisation maps}

\begin{figure*}
	\includegraphics[width=1\textwidth]{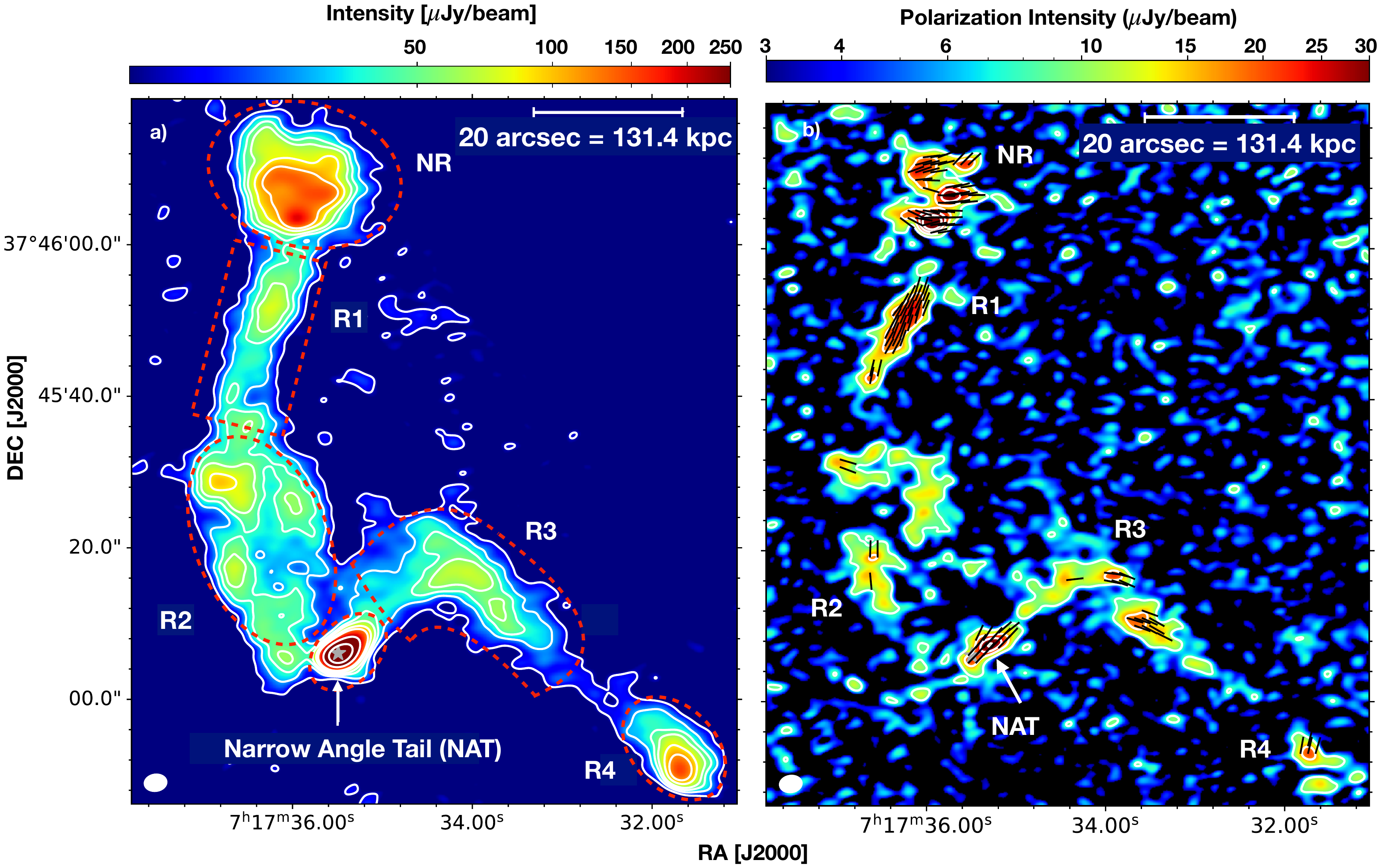}
    \caption{a) The radio relic's total intensity image at 10 GHz. The location of the NAT is indicated by an arrow, while the red dashed lines indicate the regions used for the spectropolarimetric analysis. b) The polarisation intensity map; black vectors indicating the magnetic field direction, white contours represent the polarisation intensity at [3, 6, 12] $\times$ $\sigma^{Pol}_{\text{rms}}$ levels. The size of the synthesized beam $\theta$=2$\farcs${967}$\times$2$\farcs${244} is indicated in the bottom left corner of each map.
}
    \label{fig:relicWB}
\end{figure*}

Fig.~\ref{fig:relicWB} shows the high-resolution ($\theta\sim\ $2$\farcs${96}$\times$2$\farcs${24}) total intensity (a) and the polarisation intensity (b) maps. 

The total flux density of the radio relic at this frequency is 9.5 mJy, with a sensitivity of RMS $\approx$ 5$\mu$Jy beam$^{-1}$ (1$\sigma$). The relic has a largest angular size of $\approx$ 95$\arcsec$, which at the distance of the source is equivalent to a structure of $\approx$ 620 kpc. The continuum image, Fig.~\ref{fig:relicWB}a, shows the already known complex morphology of the relic with the NAT indicated with an arrow, two arched structures at the position of R2 and R3, the elongated structure R1 and two bright regions at the very end of the relic, the northern region (NR) and R4. In the image, the six areas (NR, R1-R4 and the NAT) have been marked with dashed red lines, indicating those regions where spectropolarimetric analysis have been performed. 

At frequencies as high as 10 GHz, the radio spectrum of a radio relic could be affected by the thermal Sunyaev-Zel’dovich (SZ) effect. The SZ effect acts as a contaminant to the synchrotron flux, decreasing the observed flux from its true value. Following the discussion in \cite{Basu2016}, it is possible to estimate the contamination for our radio relic, although a precise determination would require a detailed knowledge of the projected pressure at the relic position. Following equation 14 in \cite{Basu2016} and taking into account the angular size distance (1.36$\times$$10^{3}$ Mpc) of the relic, its dimension ($\approx$ 0.620 Mpc) and assuming a value of 10$^{-3}$ cm$^{-3}$ for the thermal electron density and an approximation of T$_{X} = 20.0$ $\pm$ 4 keV for the X-ray temperature  \citep[from ][]{vanWeeren2017-835}, it is possible to estimate the total SZ flux decrement of $\approx$ $-0.56$ $\pm$ 0.21 mJy (corresponding to a contamination of $\approx$ 6\% ), which returns a true total flux of $\approx$ 10.1 mJy. However, looking at Fig. 7 in \cite{Basu2016} it is also possible to estimate the contamination from the SZ effect considering some other general radio relic properties, such as the cluster mass and redshift. This approach returns a rough approximation for the SZ contamination of $\approx$ 18\%, resulting in a true total intensity flux of 11.5 mJy, a slightly higher value than the previous one. Knowing the large uncertainties we have in all measurements, we can conclude that flux density could be underestimated by a factor of $\approx$ 6\% to 18\% due to the SZ effect. A recent thermal SZ (tSZ) analysis has only shown marginal evidence for substructures coincident with the relic \citep[see][]{Adam2018} suggesting that the flux density underestimation at 10 GHz is likely small.

A sensitivity of $\approx$ 3$\mu$Jy beam$^{-1}$, has been reached for the Stokes Q and U images; hence, emission as faint as $\approx$ 10 $\mu$Jy (3$\sigma$ detection) was detected in the polarisation intensity map, shown in Fig.~\ref{fig:relicWB}b. The polarised emission is detected at all the sub-regions with flux densities of $\approx$ 0.2-0.3 mJy for NR, R1, R2 and R3, while flux densities of $\approx$ 0.07 mJy is detected for the regions NAT and R4. The black vectors superimposed on the polarised flux density map represent the magnetic field configuration, which follows the complex morphology of the radio relic.

\begin{figure}
	\includegraphics[width=0.8\linewidth]{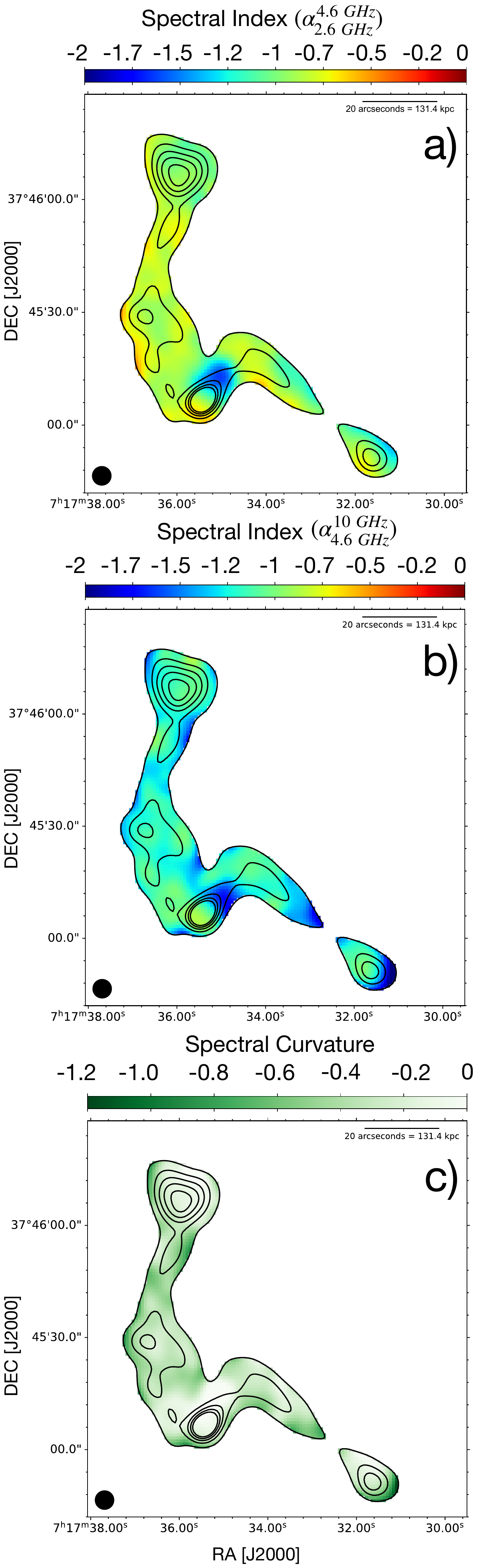}
    \caption{Spectral index maps computed between S- and C-bands a) and C- and X-bands b). Image c) shows the spectral curvature map. No strong negative curvature is detected along radio relic. The black contours are the total intensity values at [3, 6, 9, 12, 18, 24, 48, 86] $\times$ $\sigma^{I}_{\text{rms}}$ levels.}
    \label{fig:relicSpIn}
\end{figure}

Fig.~\ref{fig:relicSpIn} shows the spectral index distribution map of the relic, which has been calculated by combining high-frequency data together with low-frequency data in the S and C bands available in the literature \citep{Rajpurohit2022}. We created spectral index maps using S- and C-bands data Fig.~\ref{fig:relicSpIn}a and C- and X-bands data Fig.~\ref{fig:relicSpIn}b, where Stokes I map at X band has been corrected for a contamination of 18\% due to the SZ effect. The calculation of the spectral index has been performed assuming a spectral power-law variation of the flux ($S_\nu$) with frequency ($\nu$) in the form $S_\nu \propto \nu^\alpha$. We also derived the spectral curvature map, following the procedure presented in \cite{Rajpurohit2022} (equations 4 and 5), to investigate whether the relic shows any sign of curvature, Fig.~\ref{fig:relicSpIn}c. The maps show typical steep spectral index values in the range of the spectral index measurements $\approx -0.8$ to $-1.0$ at all the frequency coverage. The errors of the measurements are all below $\approx$ 0.2. Since there are no strong changes in the continuum spectrum, the spectral curvature map does not show any sign of a strong negative curvature with values around $\sim$ $-0.3$ in the regions where strong continuum emission is detected (see contours in Fig.~\ref{fig:relicSpIn}c).

\subsection{Stokes QU-fitting of selected regions: S- to X-bands }
\begin{figure*}
	\includegraphics[width=1\textwidth]{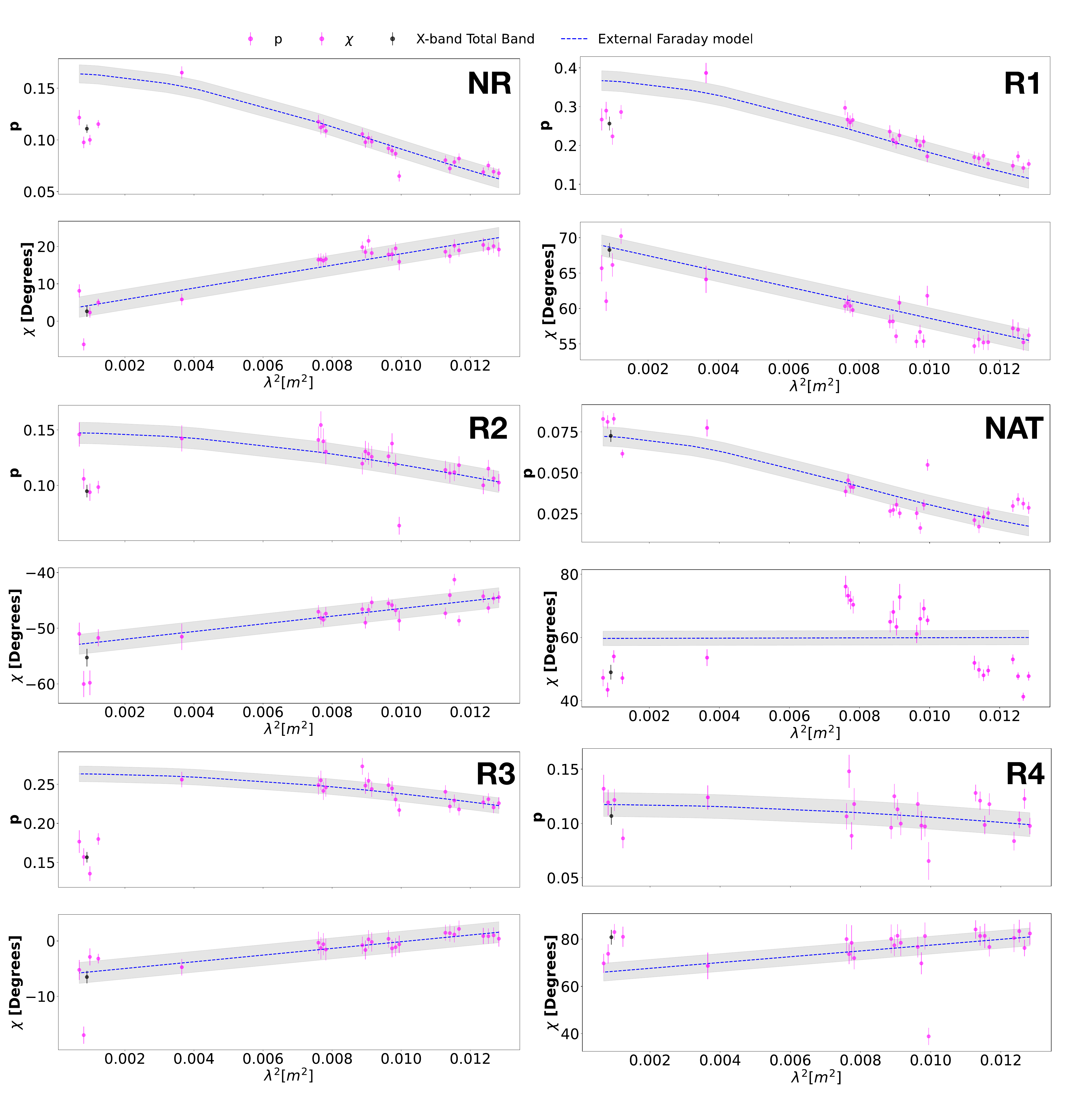}
    \caption{Result of the Stokes QU-fitting assuming one external Faraday screen for the 6 regions. All points correspond to the S-, C-, and X-bands. For each plot, the upper panel shows the fractional polarisation ($p$) and the lower panel shows the polarisation angle ($\chi$), both versus $\lambda^{2}$ for the six regions: NR, R1, R2, R3, R4 and NAT. Blue dashed lines is the fit to one external depolarisation model (equation \eqrefpar{e4.7}) with its corresponding 3$\sigma$ error.}
    \label{fig2}
\end{figure*}

\begin{figure*}
	\includegraphics[width=1\textwidth]{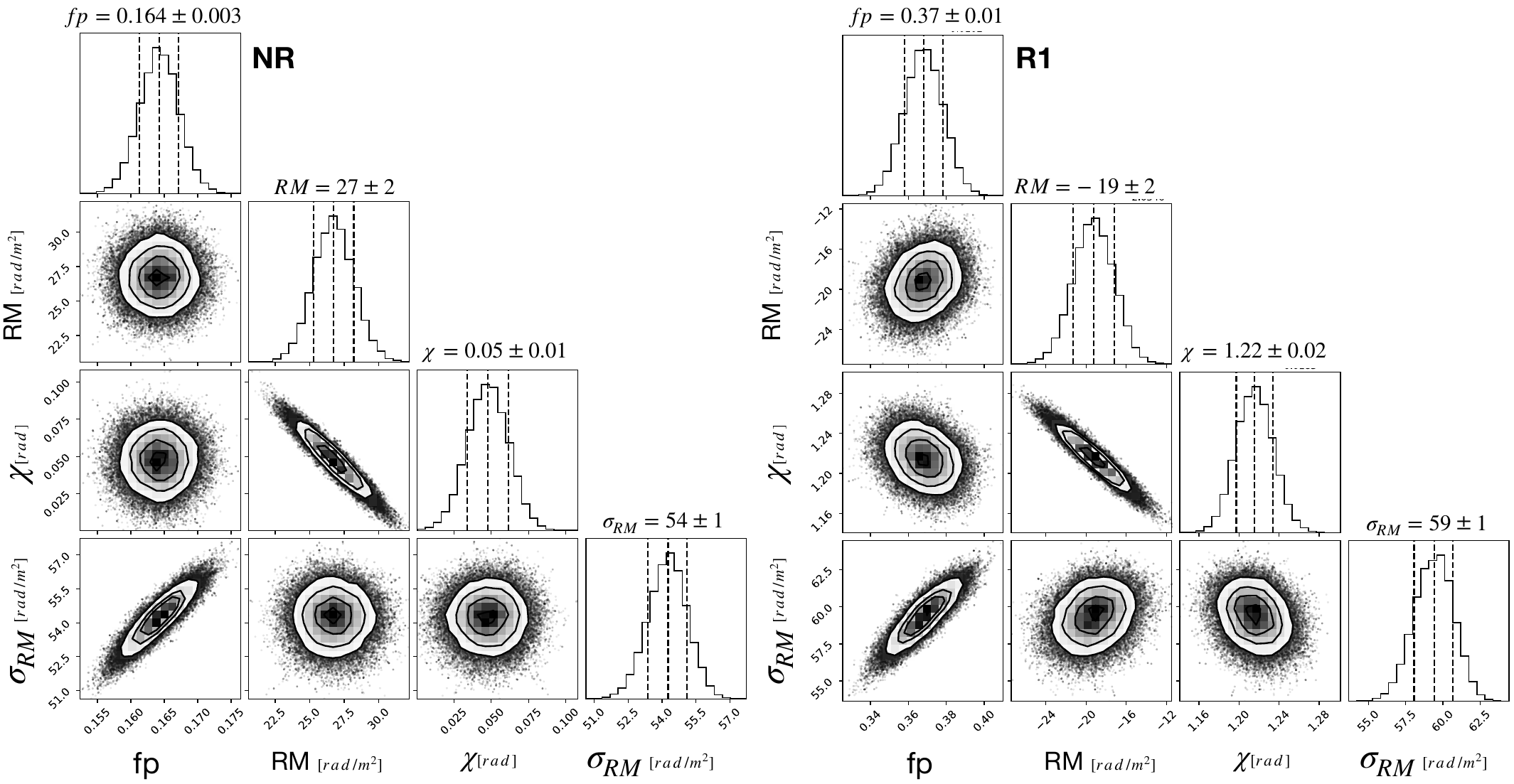}
    \caption{Corner plots of the Stokes QU-fitting modelling (one external Faraday screen) for the NR and R1 regions.}
    \label{fig:mcmc1}
\end{figure*}

\begin{figure*}
	\includegraphics[width=1\textwidth]{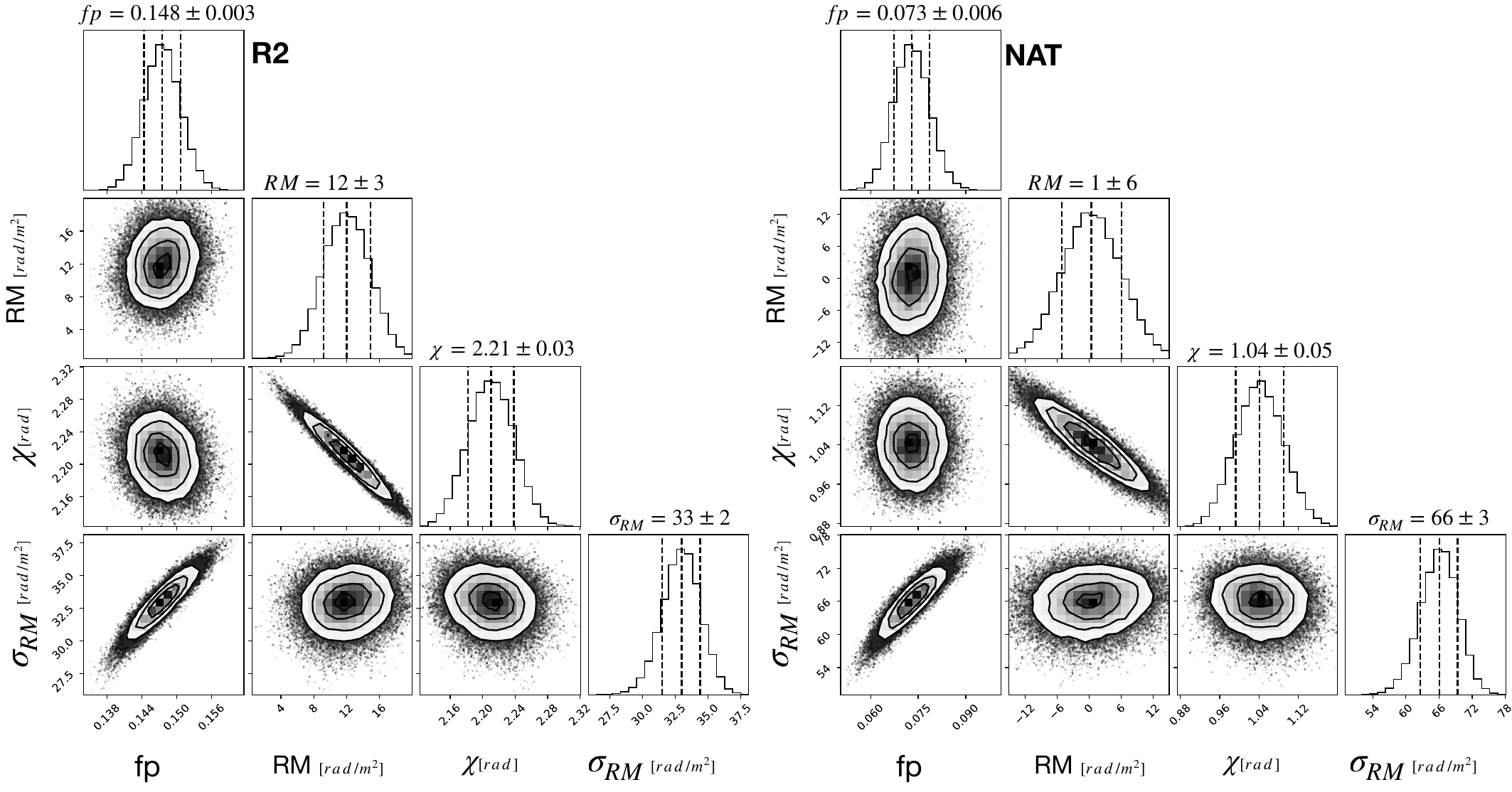}
    \caption{Corner plots of the Stokes QU-fitting modelling (one external Faraday screen) for the R2 and the NAT regions.}
    \label{fig:mcmc2}
\end{figure*}

\begin{figure*}
	\includegraphics[width=1\textwidth]{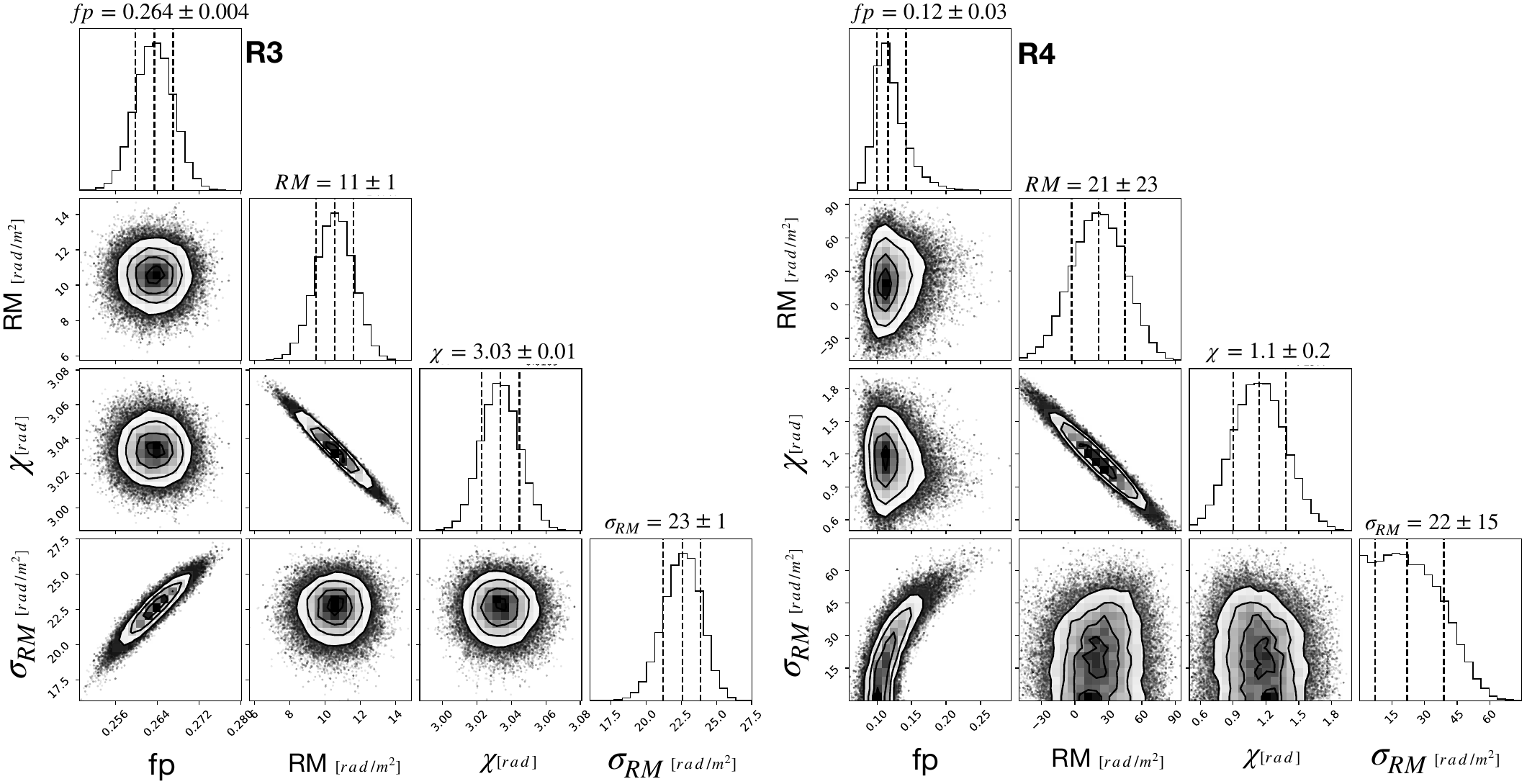}
    \caption{Corner plots of the Stokes QU-fitting modelling (one external Faraday screen) for the R3 and R4 regions.}
    \label{fig:mcmc3}
\end{figure*}

 Stokes QU-fitting has been performed in six areas along the relic (NR, R1-R4 and NAT). To understand the behaviour of the polarised emission of the relic, the total X-band data, which was split into four datasets each of 1GHz-wide, and the S- and C-band data, available from the literature \citep{Rajpurohit2022}, have been added to the analysis. The Stokes I, Q, and U parameters have been extracted from the areas of interest along the wide frequency coverage, after convolving all images to the circular beam of $\theta$= \(5''\). An external Faraday screen, an internal Faraday screen and a combination of the two have been fitted. It is worth noting that the equations we are using for the model fit are assuming a single value for the RM, for the intrinsic fractional polarisation (\textit{p}), and for the intrinsic polarisation angle (\textit{$\chi_{0}$}) across the region to fit and that previous low-frequency work indicates a variation of the the polarisation properties with the distance from the shock front using rectangular regions of width 2\arcsec \citep[see Fig.4 in ][]{Rajpurohit2022}. For that study, the authors had enough sensitivity to perform such analysis. Our 10 GHz data do not have enough sensitivity in polarisation to recover the same polarised structures detected at low-frequency. However, we extracted the average value of the Stokes parameters I, U, and Q in those regions where the 10 GHz data show polarisation signal and we analyzed them together with the average values of the Stokes parameters I, U, and Q at low-frequency extracted from the same area. We are aware that future high-frequency and high-sensitivity data will be vital for a better analysis. For all six regions, the best representation of all the data is with an external Faraday screen, as shown in Fig. \ref{fig2}. The plots show the behaviour of the observed fractional polarisation $p$($\lambda^{2}$) (upper panel) and the polarisation angle (lower panel) with wavelength $\lambda^{2}[m^2]$, the model fit (blue dashed line), corresponding to the equation \eqrefpar{e4.7}, and the $3\sigma_{\text{rms}}$ error derived from the MCMC fit (gray band). 
The MCMC uncertainties are shown in corner plots in Figs. \ref{fig:mcmc1}, \ref{fig:mcmc2}, and \ref{fig:mcmc3}. 

It is clearly seen that the fractional polarisation data at 10 GHz do not follow the behaviour of an external Faraday model at all, especially for the regions NR, R1, R2 and R3. The X-band data seem to trace a sinusoidal behaviour rather than following the exponential trend of the low-frequency data points. As seen in the plots of Fig.~\ref{fig2} the gray area, representing the $3\sigma$ error bar of the fit, is not matching the X-band data, which show a different distribution of the polarisation parameters. The high-frequency, polarised data have a low SNR, therefore, the model fit is not capable of anchoring a good intrinsic value for both the fractional polarisation and the polarisation angle. To increase the SNR of the 10 GHz data, the total band Stokes I, Q, and U parameters have been extracted and added as black data points in the plots of Fig.~\ref{fig2}. Considering these wide-band data points, the fit still does not describe the complexity of the relic's polarisation properties. In the next section, the polarisation properties, using the 10 GHz data only, are investigated in order to explore whether the high and the low frequencies are tracing different regions/behaviours.


\subsection{Depolarisation modelling on the 10 GHz maps}
\begin{figure*}
	\includegraphics[width=1\textwidth]{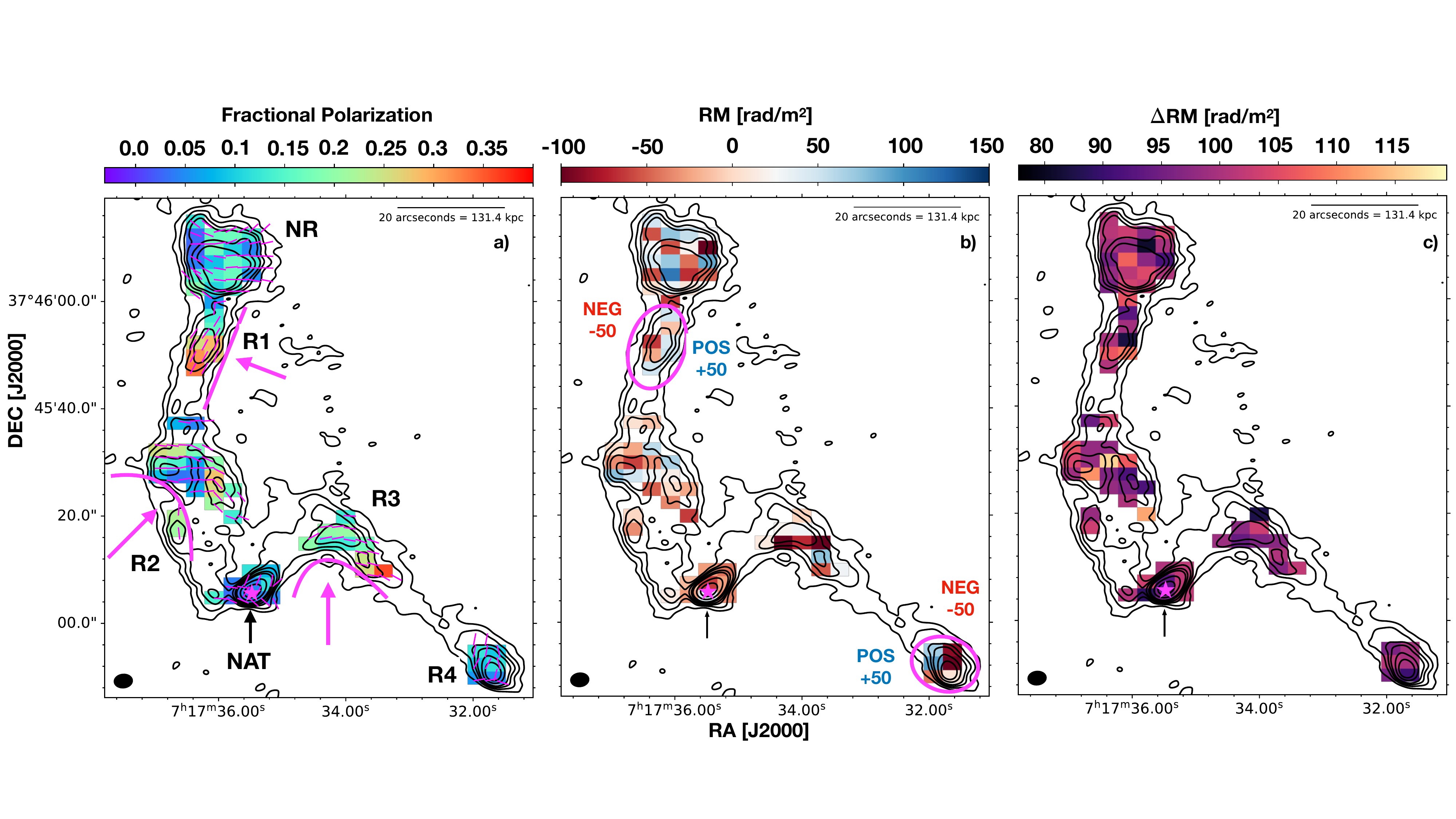}
    \caption{a) Intrinsic fractional polarisation map with the orientation of the magnetic field as magenta vectors b) Rotation measure (RM) map. c) $\Delta_{\text{RM}}$ map. The three maps are the result of an internal Faraday depolarisation model. The NAT is indicated by a black arrow; in figure a) the pink lines with pink arrows indicate the approximate location of the regions where shock fronts are supposed to be present; in figure b) the pink solid lines, mark the two regions where a change of the RM sign is visible. In all the images, the black contours are derived from the total intensity map at [3,6,9,12,18,24,48,86] $\times$ $\sigma^{I}_{rms}$ levels. The size of the synthesized beam is indicated in the bottom left corner of each map.}
    \label{fig:Fig3}
\end{figure*}

To better understand the high-frequency polarisation behaviour, Stokes QU-fitting has been performed using only the high-frequency Stokes I, Q, and U images, each averaging 1GHz of its wide-band and all convolved to the same angular resolution of $\theta$ = 3$\farcs${500}$\times$2$\farcs${642}. The fitting has been performed by analyzing these maps using an area equal to the convolved beam size. External and internal Faraday screen models have been used for the analysis. Because of the poor sampling of the data, only four data points, the spectropolarimetric analysis at high-frequency did not return a preferred model fit. However, when observing the distribution of the intrinsic external and internal polarisation angle, it is possible to see that an external Faraday depolarisation model returns a more randomly oriented intrinsic magnetic field configuration, while the internal Faraday depolarisation model returns a much more ordered magnetic field, following the different substructures of the relic. Considering that the six regions showing polarised signal are quite large structures (approximately 60 kpc and 100 kpc in size) and given the low SNR of the available data, it is unlikely that we can achieve sufficient accuracy to detect strong variations in the magnetic-field configuration at the scale of individual beam cells. We are aware that this visual inspection is not decisive to determine which is the representative model, still we are more inclined to prefer an internal Faraday depolarisation as the mechanism describing the high-frequency data. The results of the internal Faraday depolarisation are shown in Fig.~\ref{fig:Fig3}, where the intrinsic fractional polarisation with superimposed magnetic field line configuration (a), the RM distribution (b), and the $\Delta_{\text{RM}}$ (c) maps are reported. The results of the external Faraday screen model are shown in Appendix, Fig. \ref{AppExternal}.

The intrinsic fractional polarisation map varies with values ranging from $\sim$10$\%$ to $\sim$ 30$\%$ across the radio relic (mean value within the beam area). Larger values of intrinsic fractional polarisation can be seen in the regions of the predicted shock fronts R1, R2, and R3 (marked with pink lines and with the pink arrows marking the direction of the shock fronts, Fig.~\ref{fig:Fig3}a) where typical values are $\sim20-30\%$. These regions also show a well-ordered magnetic field configuration (magenta lines superimposed), following the morphology of the relic. An increase in fractional polarisation values is supported by HDM simulations, which demonstrate that shock compression of a randomly oriented magnetic field results in a magnetic field alignment and, consequently, a high polarisation fraction \citep{DominguezFernandez2021}. The regions with a lower intrinsic fractional polarisation, such as NR, NAT, and R4, with values around $\sim10\%$, indicate a less homogeneous medium with a more complex magnetic field configuration, which might cause stronger depolarisation. 

The Galactic foreground RM at the position of the cluster region exhibits a relatively small mean value \citep[i.e., $\sim$ -6 rad/m$^2$,][]{Hutschenreuter2020} compared to the values obtained here. Consequently, the presented measurements are not dominated by the Galactic RM contribution. Fig.~\ref{fig:Fig3}b shows the RM map (observed) above the 3$\sigma$ level detection. To estimate the RM in the rest-frame of the source, the following correction to the redshift $\textit{z}$ is needed:

\begin{equation}
 RM_{RF} = RM_{obs}(1+z)^{2}
    \label{RM_restframe}
\end{equation}

\noindent where RM$_{RF}$ is the RM in the rest-frame and RM$_{obs}$ is the observed RM, which results from modelling. The RM map shows several substructures. The NAT core shows homogeneous values of $\approx$25 rad/m$^2$, which correspond to $\approx$60 rad/m$^2$ at the redshift of the source, while R1 and R4 show higher values and some interesting gradients. Both regions show values of $\approx$50 rad/m$^2$, corresponding to $\approx$120 rad/m$^2$ at the redshift of the source, but with a change in the sign of the RM from positive to negative and highlighted with pink ellipses in Fig.~\ref{fig:Fig3}b to help visualize them. A gradient of the RM may indicate a sign of a rotation of the magnetic field configuration, most likely because of shock compressions and/or surrounding medium interactions. The two arched structures R2 and R3 show predominantly negative RM values, again around $\approx$-50 rad/m$^2$, corresponding to $\approx$-120 rad/m$^2$ at the source redshift. A more chaotic RM distribution is shown at the position of NR, suggesting that the magnetic field structure of the relic is complex and is influenced by possible interaction with the surrounding medium.

The $\Delta_{\text{RM}}$ map presented in Fig.~\ref{fig:Fig3}c reveals values along the relic in the range of $\approx$95-100 rad/m$^2$, which, at the redshift of the target, is $\approx$230-240 rad/m$^2$. The $\Delta_{\text{RM}}$ basically indicates how smoothly the magnetic field is changing within the observed beam. This map does not show strong variations, meaning that an ordered change of the magnetic field configuration across different lines of sight through the relic, is occurring.

\section{Discussion}
\label{sec:5}
The results presented in this work suggest a change in the depolarisation mechanism when the relic is observed at high-frequency. As seen in the plots of Fig.~\ref{fig2}, the low-frequency S- and C-bands are better described with an external Faraday depolarisation mechanism. This trend is in agreement with previous low-frequency work \citep[][]{Rajpurohit2022}, where the high-sensitivity data allow the authors to fit up to two external Faraday components. Here, due to the low SNR of the data at 10 GHz and because we are averaging the Stokes parameters for large regions, we are limited to look for trends in the data. At frequencies higher than C-band, no external depolarisation should be present, and we should already be detecting the intrinsic polarisation fraction. However, for all six regions, the emission at X-band is not following the expected polarisation behaviour. The high-frequency X-band data behaves in a different way with respect to the low-frequency data (see Fig.~\ref{fig:Fig3}). Moreover, the regions NR, R1, R2 and R3 clearly show that the mean value of the observed fractional polarisation at X-band (black points in Fig.~\ref{fig2}) is systematically lower than the values calculated from the C-band images, meaning a stronger depolarisation is affecting the high-frequency data. These data are, most likely, tracing different synchrotron emission region with different physical properties. The analysis of the X band data only does not suggest a strong preference for either models. However, these findings point towards an internal depolarisation mechanism because the magnetic field configuration resulting from the internal Faraday depolarisation model is much more ordered than that resulted from modelling an external Faraday layer.

In the following, we explore several scenarios to explain what is causing the drastic change in the observed fractional polarisation. One of these scenarios is an internal depolarisation mechanism at these physical scales. Albeit we acknowledge that for a proper clarification on this matter, high-frequency and high-sensitivity radio data are necessary. 

We investigated whether the observed fractional polarisation drop could be associated with the extended emission resolved out at X-band because of its angular resolution. That is very unlikely since the VLA X-band receiver at C array is sensitive to a scale as large as 145$\arcsec$ and the largest scale of the relics is $\approx$ 90$\arcsec$. Moreover, when performing the depolarisation analysis we convolved all the images to the same angular resolution, and we normalized the polarisation flux density at all the frequency bands with their corresponding Stokes I flux densities.

We also investigated whether such strong depolarisation at X-band could be due to in-band depolarisation since we are averaging the Stokes I, Q and U data within 1GHz bandwidth. However, this is also very unlikely because such depolarisation would imply the presence of very high RM with values larger than 5000 rad/m${^2}$, which is unexpected for a radio relic.

We finally investigate the scenario in which the 10 GHz spectropolarimetric analysis is described by an internal Faraday depolarisation. We have seen that the observed 10 GHz polarisation data do not follow the external Faraday depolarisation mechanism (see the plots of Fig.~\ref{fig2}). Indeed, the external Faraday depolarisation mechanism predicts a constant value towards the higher frequency. Instead, the observed data do not follow that prediction. When the high-frequency data were analyzed alone, they did not suggest any preferred model to follow. However, we tend to prefer an internal scenario because the distribution of the intrinsic magnetic field configuration lines is more ordered with respect to that resulting from an external layer (see maps in Fig.~\ref{fig:Fig3}). Because higher-frequency radio emission traces regions closer to the shock front, and lower-frequency emission is dominated by downstream plasma, as we observe from low to high frequencies we are tracing synchrotron-emitting regions which are characterized by increasingly more energetic particles and we are tracing deeper layers. The continuum emission at 10 GHz is tracing more compact synchrotron structures than those at the lower frequencies, therefore, their intrinsic polarisation properties, such as the intrinsic magnetic field structure and the local plasma density fluctuations, could play a more significant role. While at lower frequencies, the dominant effect is an external depolarisation with its exponential decay starting at C-band and being stronger at S-band, at higher frequencies the polarised data are describing a completely different mechanism most likely dominated by the intrinsic magneto-ionic properties of those more compact structures. 

Moreover, the presented intrinsic X-band properties are consistent with theoretical predictions. The observed $\Delta_{\text{RM}}$ ranging $\approx$95-100 rad m$^{-2}$ aligns with numerical simulations by \citet{DominguezFernandez2021}, who modeled internal RM dispersion within a (200 kpc)$^3$ volume. Their work demonstrates that $\Delta_{\text{RM}} < 100$ rad m$^{-2}$ corresponds to early stages of the shock propagation, assuming a magnetic field strength of 1.5 $\mu$G. This agreement suggests our measured dispersion reflects internal dispersion at similar physical scales.

Furthermore, our findings show consistency with observational studies of similar systems. \citet{Stuardi2019} reported comparable $\Delta_{\text{RM}} \sim 100$ rad m$^{-2}$ in the western relic of RXC J1314.4-2515, attributing this to internal Faraday rotation. 

The differences observed in the X-band data may provide critical insights into the internal magnetic field configuration of the source. The presence of an internal Faraday screen component would imply a more intricate magnetized plasma environment with the presence of thermal and non-thermal electrons, which may provide a deeper understanding of the nature of the (re-)accelerated particles in radio relic systems. It is worth to clarify that it is still a matter of debate whether the non-thermal electrons (re-)accelerated and forming the radio relics are thermal electron from the ISM or non-thermal electrons previously ejected from an AGN. In this study we conclude that a possible scenario explaining the changes in polarisation could be due to an internal depolarisation. For such internal depolarisation model, the source that rotates the
angle and depolarizes is the same synchrotron source, i.e., the non-thermal electrons ejected from the AGN jet at the centre of the radio relic structure, for instance.

To deepen our understanding of this topic, it is important, when possible, to explore these giant structures at high-frequency, in order to trace the continuum emission and their intrinsic polarisation properties of the most compact structures, and to connect them with the appropriate particle acceleration mechanisms. High-frequency polarisation observations are the best option for examining the synchrotron emission of the most compact components, as they can effectively trace the shock fronts and investigate whether those regions are better described with the AGN magnetic field and/or the large scale relic magnetic field. It is important to perform deep observations (reaching the $\mu$Jy beam$^{-1}$) of the radio relics at high frequencies, such as X, K and Q bands, since these sources are optically thin at these frequencies and these bands are those sensitive to the radio emission of the closest region of a shock front, where compression of the plasma is occurring. Moreover, radio relics are quite strong sources of polarisation therefore, one can estimate at least $\sim$20\% of the continuum flux to be polarised, which is quite large. Performing spectropolarimetric studies at high-frequency gives the opportunity to investigate the intrinsic magneto-ionic structures of a shock region and deepen the knowledge on the (re)-acceleration mechanism.

\section{Summary and conclusions}
\label{sec:6}

In this work, a spectropolarimetric analysis at high-frequency (X-band) has been conducted on the radio relic of the merging galaxy cluster MACS J0717.5$+$3745. Moreover, data from the literature at S- and C-bands have been added to perform a wide frequency coverage Stokes QU-fitting analysis with the aim of describing the depolarisation mechanism affecting the source. 

The results presented in this work suggest a complete different behaviour of the polarised data at high-frequency from those at low-frequency. The high-frequency X-band data are most likely tracing an internal magnetized medium that is responsible for the complex behaviour of the polarised data.% This internal medium is thought to be filled with relativistic electrons previously emitted from the NAT source and successively accelerated by the shocks which are present within the cluster. 
Moreover, the fractional polarisation emission from the more compact X-band structures is systematically lower than the value predicted by the external modelling therefore, a stronger depolarisation is occurring at this frequency range. As we observe at higher frequencies we are observing increasingly more energetic particles and we are tracing deeper layers, the intrinsic magneto-ionic properties of which are affecting the emission differently with respect to those at low-frequency.

We suggest that the structure and the properties of the magnetic field in the radio relic may be more complex than currently understood from observations at the lower frequency bands. These findings contribute to a deeper understanding of the underlying physical processes within these complex radio relics. Not only this, these results underline the importance to perform, when possible, high-frequency observations as they can effectively trace the shock fronts and trace the intrinsic properties of the high-energy electron population. This allows for the characterization of their intrinsic magneto-ionic structure and a more thorough analysis of their true nature.

Future facilitates, such as the planned next generation VLA (ngVLA), will be able to achieve incredible sensitivity (sub-$\mu$Jy beam$^{-1}$) and great angular resolution to detect and trace all the substructures involved in the shocks of radio relics.

%We interpreted the sudden rise of the fractional polarisation at C band with the effect of a repolarisation, i.e. the fractional polarisation value increases towards the lower frequency. The 10 GHz fractional polarisation, which suffers from an internal depolarisation at high-frequency, is repolarised when it passes through a new and more extended magneto-ionic component with a relatively ordered magnetic field, completely disconnected from the high-frequency depolarised layer. This encounter contributes to an increase of the observed linear polarisation at these frequencies. Moving towards the lower S-band frequency range, the polarised data follow an external Faraday screen.

\section*{Acknowledgements}
A.P. acknowledges support from UNAM DGAPA-PAPIIT grant  IA100425; this work benefited from the UNAM-NRAO Memorandum of Understanding in the framework of the ngVLA Project (MOU-UNAM-NRAO-2023). A.P. acknowledges Carlos Carrasco-Gonz\'alez for his fruitful comments and suggestions.
E.F.-J.A. acknowledges support from UNAM- PAPIIT projects IA102023 and IA104725, and from CONAHCyT Ciencia de Frontera project ID: CF-2023-I- 506.
We are grateful for the valuable feedback from the anonymous reviewer(s), whose input strengthened the paper.

%%%%%%%%%%%%%%%%%%%%%%%%%%%%%%%%%%%%%%%%%%%%%%%%%%
\section*{Data Availability}
The data underlying this article will be shared on reasonable request to the corresponding author.

%%%%%%%%%%%%%%%%%%%% REFERENCES %%%%%%%%%%%%%%%%%%

% The best way to enter references is to use BibTeX:

\bibliographystyle{mnras}
\bibliography{example} % if your bibtex file is called example.bib

% Alternatively you could enter them by hand, like this:
% This method is tedious and prone to error if you have lots of references
%\begin{thebibliography}{99}
%\bibitem[\protect\citeauthoryear{Author}{2012}]{Author2012}
%Author A.~N., 2013, Journal of Improbable Astronomy, 1, 1
%\bibitem[\protect\citeauthoryear{Others}{2013}]{Others2013}
%Others S., 2012, Journal of Interesting Stuff, 17, 198
%\end{thebibliography}

%%%%%%%%%%%%%%%%%%%%%%%%%%%%%%%%%%%%%%%%%%%%%%%%%%

%%%%%%%%%%%%%%%%% APPENDICES %%%%%%%%%%%%%%%%%%%%%
\newpage
\appendix
 
\section{Depolarisation modelling}
\label{sec:appendix A} 
\begin{figure*}
	\includegraphics[width=1\textwidth]{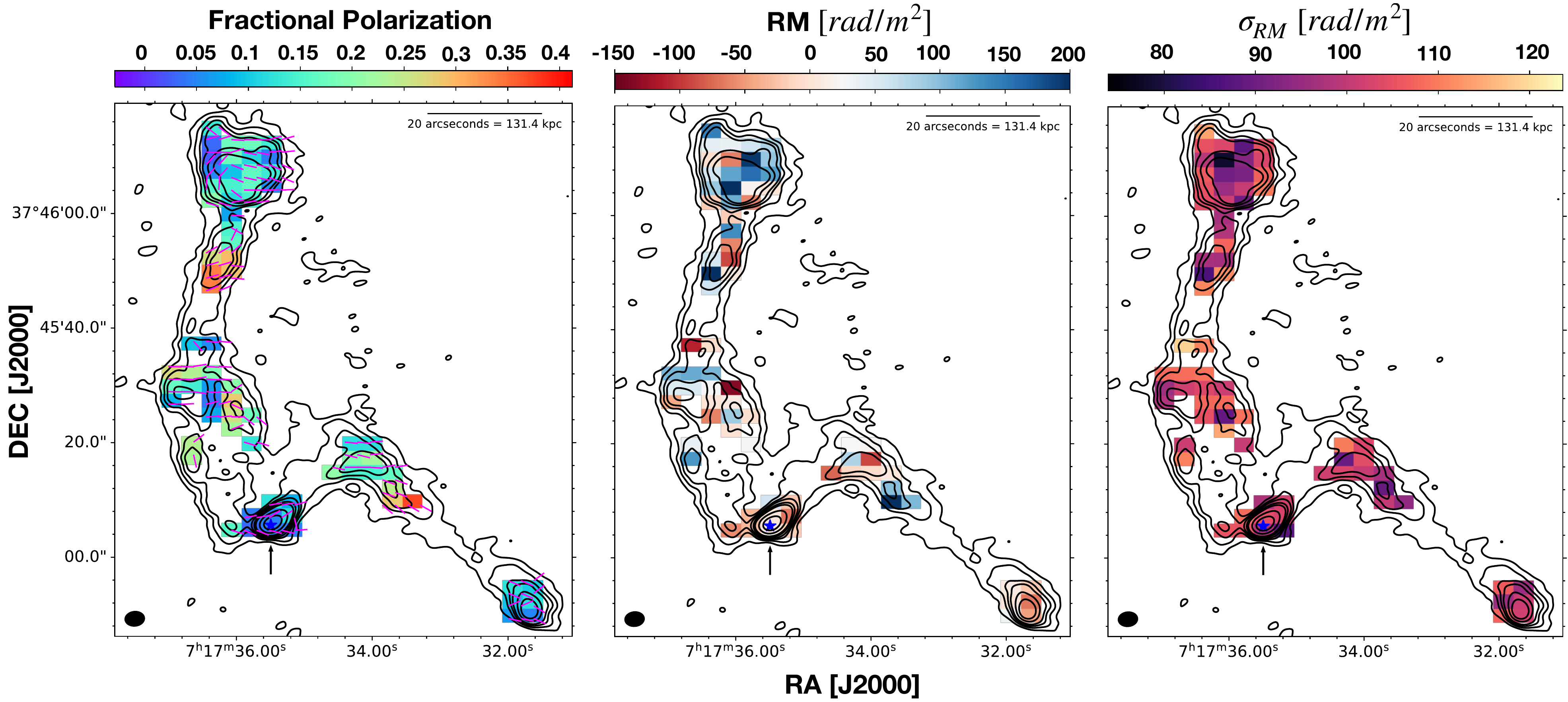}
    \caption{a) Intrinsic fractional polarisation map with the orientation of the magnetic field as magenta vectors b) Rotation measure (RM) map. c) $\sigma_{\text{RM}}$ map. The three maps are the result of the external Faraday depolarisation model.}
    \label{AppExternal}
\end{figure*}

%\begin{figure*}
%	\includegraphics[width=1\textwidth]{NUEVA_REGION_NR_modelos.pdf}
%    \caption{example of depolarisation modelling for the region NR where we performed one and two internal Faraday depolarizing layers and one and two external Faraday depolarizing layers. The X band data do not follow non of the models.}
%    \label{TESTModels}
%\end{figure*}
%%%%%%%%%%%%%%%%%%%%%%%%%%%%%%%%%%%%%%%%%%%%%%%%%%

% Don't change these lines
\bsp	% typesetting comment
\label{lastpage}
\end{document}